\documentclass[aip,jcp,reprint]{revtex4-1}
\usepackage{graphicx,xcolor,soul}
\usepackage{amsfonts,amssymb,amsmath,bm}

\definecolor{Red}{rgb}{0.832, 0.367, 0}

\newcommand{\icp}{\varphi}
\newcommand{\ivf}{\psi}
\newcommand{\extfield}{\phi_{\text{ext}}}
\newcommand{\kB}{k_\text{B}}
\newcommand{\tA}{\tilde{A}}

\newcommand{\rmd}{\textrm{d}}
\newcommand{\va}[1]{\bm{a}_{#1}}
\newcommand{\vecr}[1]{\bm{r}_{#1}}
\newcommand{\vx}{\bm{x}}

\newcommand{\spd}{\rho}
\newcommand{\tpd}{\rho^{(2)}}
\newcommand{\fix}{\text{fix}}
\newcommand{\trial}{\text{tr}}

\begin{document}

\title{Free-Energy Functional Method for Inverse Problem of Self Assembly} 


\author{Masashi Torikai}%
\email[]{torikai@phen.mie-u.ac.jp}
\affiliation{Department of Physics Engineering, Faculty of Engineering, Mie University, 1577 Kurimamachiya-cho, Tsu, Mie 514-8507, Japan}


\date{9 April 2015}

\begin{abstract}
A new theoretical approach is described for the inverse self-assembly problem, i.e., the reconstruction of the interparticle interaction from a given structure.
This theory is based on the variational principle for the functional that is constructed from a free energy functional in combination with Percus's approach [J. Percus, Phys. Rev. Lett. \textbf{8}, 462 (1962)].
In this theory, the interparticle interaction potential for the given structure is obtained as the function that maximizes the functional.
As test cases, the interparticle potentials for two-dimensional crystals, such as square, honeycomb, and kagome lattices, are predicted by this theory.
The formation of each target lattice from an initial random particle configuration in Monte Carlo simulations with the predicted interparticle interaction indicates that the theory is successfully applied to the test cases.
\end{abstract}

\pacs{}

\maketitle 

\section{Introduction}\label{sec:1}
In condensed phases, diverse interactions between the constituents (such as atoms, molecules, micelles, nano- and microparticles) produce a variety of simple and complex structures.
Besides a plethora of the crystal, liquid crystal, and quasicrystal structures formed by conventional atoms and molecules, square and honeycomb lattices of colloidal nanocrystals,~\cite{Evers2013} kagome lattice of triblock Janus particles,~\cite{Chen2011} and quasicrystals of dendrimers~\cite{Zeng2004} and binary nanocrystals~\cite{Talapin2009} are additional experimental examples of non-trivial structures.
Molecular simulations for systems with several interparticle interactions have been performed to enumerate the stable structures;
it has been found that particles with a rather simple interaction potential can assemble into complex structures, e.g., hard sphere (HS) plus linear ramp,~\cite{Jagla1998} HS plus square well,~\cite{Skibinsky1999} HS plus square shoulder,~\cite{Dotera2014} and Lennard-Jones plus Gaussian~\cite{Engel2007} (LJG) potential particles assemble into a two-dimensional (2D) quasicrystal.
Density functional theory (DFT), another tool to find the thermodynamically stable phases, has shown that in three dimensions (3D), the LJG fluid forms a body centered cubic (bcc) crystal as a stable phase.~\cite{Suematsu2014}

When a structure of a condensed phase is known from experiments or is designed artificially, determination of the interparticle interaction is more efficient by using the inverse approach, in which the interparticle interaction is derived from the given structure, than using the forward approach of exhaustive search by the molecular simulations or DFT.
The inverse statistical mechanical method,~\cite{Rechtsman2005} which is the most elaborated inverse approach, has been successfully used to design the interparticle interaction potentials that generate the square, honeycomb,~\cite{Rechtsman2005,Rechtsman2006,Marcotte2011,Jain2014} kagome, and rectangular~\cite{Zhang2013,Jain2014} lattices in 2D, and the simple cubic, bcc, simple hexagonal,~\cite{Rechtsman2006a,Jain2013b,Jain2014} diamond,~\cite{Rechtsman2007,Jain2013b,Jain2014} and wurtzite lattices~\cite{Rechtsman2007} in 3D.
In the inverse statistical mechanical method, the interparticle interaction is optimized with respect to the energy and mechanical stability criteria via molecular simulations.
The studies on the inverse problem at finite temperature have mostly been based on the molecular simulations at finite temperature.
(It was suggested in Ref.~\onlinecite{Jain2013a} that the molecular simulations at finite temperature may not be necessary.
This is based on the observation that the target structure shows good stability at finite temperature if the interaction potential is optimized at zero temperature via the minimization of a specific simulated annealing energy defined in Ref.~\onlinecite{Jain2013b}.
However, this strategy is also dependent on computer simulations.)
For the liquid target phases, the reverse Monte Carlo (MC) method,~\cite{McGreevy1988,Lyubartsev1995} which also uses the MC simulation in the optimization, is another successful inverse approach.

Here the aim is to formulate a theory to reconstruct the interparticle potential that can stabilize a given target structure (more specifically, the single- and two-particle distributions) at finite temperature.
In this paper, I present a new simulation-free inverse method that is a variational method based on the free-energy functional theory~\cite{Caillol2002} akin to DFT.
The interparticle interaction function is defined as the function that gives the maximum of the functional.
I applied this method to the square, honeycomb, and kagome lattices, and obtained the corresponding interparticle potentials.
The potentials are then used in a series of simulated annealing MC calculations starting from random configurations.
In most cases, the resulting solid contains a few grain boundaries and many defects.
However, it is found that for each predicted potential, in at least a few percent of the simulations the particles spontaneously form the appropriate target lattice with a small number of defects.
Although the small success rate in the simulations indicates that the interaction potentials obtained here are not entirely optimized as the potentials obtained by the inverse statistical mechanical method, the observed self-assembly into the target lattice implies that the method introduced here is another promising approach to the inverse problem of the self-assembly.

\section{Theory}\label{sec:2}
We consider a single component system in a $d$-dimensional space.
The system comprises particles interacting with a pairwise-additive potential $v(\vx)$.
The grand potential of the system, with the temperature $T$, chemical potential $\mu$, and volume $V$, in the presence of an external field $\extfield(\vx)$, is defined as
\begin{equation}
 \Omega[\icp] = - \beta^{-1} \ln \Xi[\icp],
\end{equation}
where $\beta = 1/k_{\text{B}} T$ is the inverse temperature, $\icp(\vx) = \mu - \extfield(\vx)$ is the intrinsic chemical potential, and $\Xi[\icp]$ is the grand partition function
\begin{align}
 \Xi[\icp] &= \sum_{N=0}^{\infty} \int \frac{\rmd \vecr{}^{(N)}}{\lambda^{dN}N!}  \notag \\
 &\quad \times \exp\biggl[-\beta \sum_{i}^{N-1}\sum_{j>i}^{N} v(\vecr{i} - \vecr{j}) + \beta \sum_{i}^{N} \icp(\vecr{i})\biggr],\label{eq:grand_partition_function}
\end{align}
where $\vecr{i}$ and $\lambda$ denote the coordinate of the $i$th particle and the de Broglie thermal wavelength, respectively.

The key functional for this approach is
\begin{equation}
 \tA[\spd, \ivf] = \Omega[\ivf] + \int \spd(\vx, [\icp]) \ivf(\vx) \rmd \vx, \label{eq:tA}
\end{equation}
where $\spd(\vx, [\icp])$ denotes the single-particle density in the presence of the intrinsic chemical potential $\icp(\vx)$ and $\ivf(\vx)$ is an independent function.
The maximum of $\tA[\spd, \ivf]$ with respect to $\ivf$, i.e., 
\begin{equation}
 A[\spd] = \sup_{\ivf} \tA[\spd, \ivf], 
\end{equation}
is the intrinsic free energy, the Legendre transformation of the grand potential $\Omega[\ivf]$.
The maximum of $\tA[\spd, \ivf]$ is achieved when $\ivf$ is equivalent to the intrinsic chemical potential $\icp$, i.e., $A[\spd] = \tA[\spd, \icp]$.
The inequality relation 
\begin{equation}
 \tA[\spd, \icp] \ge \tA[\spd, \ivf] \quad \text{for any $\ivf(\vx)$} 
\end{equation}
provides us the variational method~\cite{Caillol2002} to determine the intrinsic chemical potential $\icp(\vx)$ that gives rise to a given density profile $\spd(\vx, [\icp])$: the $\psi(\vx)$ that maximizes the functional $\tA[\spd, \ivf]$ is the intrinsic chemical potential $\icp(\vx)$.
The variational method for $\tA[\spd, \psi]$, in which the proper $\icp(\vx)$ is obtained for a given $\spd(\vx)$, can be viewed as the inverse of the DFT,~\cite{Caillol2002, Hansen2013} in which the proper $\spd(\vx)$ is obtained for a given $\icp(\vx)$.

The variational method for $\tA[\spd, \psi]$ can be adapted to obtain the information about the interaction potential $v(\vx)$ using Percus's idea.~\cite{Percus1962, Hansen2013}
If a particle is fixed at the origin of the coordinate system, the remaining particles feel the external field $v(\vx)$ due to the fixed particle in addition to the original external field $\extfield(\vx)$.
In this situation, the single-particle density becomes $\spd(\vx, [\icp_{\fix}])$, where $\icp_{\fix}(\vx) = \mu - [\extfield(\vx) + v(\vx)] = \icp(\vx) - v(\vx)$ is the intrinsic chemical potential in the presence of the fixed particle.
Percus~\cite{Percus1962} showed that $\spd(\vx, [\icp_{\fix}])$ is related to the single- and two-particle density in the absence of the fixed particle by
\begin{equation}
 \spd(\vx, [\icp_{\fix}]) = \frac{\tpd(\vx, 0, [\icp])}{\spd(0, [\icp])}, \label{eq:PercusIdea}
\end{equation}
where $\tpd(\vx, \vx', [\icp])$ is the two-particle density between $\vx$ and $\vx'$ in the absence of the fixed particle.
As discussed above, the function $\psi$ that maximizes $\tA[\spd(\vx, [\icp_{\fix}]), \psi]$ is $\icp_{\fix}(\vx)$.
Combining the variational method for $\tA[\spd, \psi]$ with Percus's idea, we find that the function $\psi(\vx)$ that maximizes $\tA\bigl[\tpd\big/\spd, \psi\bigr]$ is $\icp_{\fix}(\vx)$.
Thus, for a given set of $\spd(\vx)$ and $\tpd(\vx, \vx')$, we can obtain the interaction potential $v(\vx) = \icp(\vx) - \icp_{\fix}(\vx)$ through the maximization of $\tA\bigl[\tpd\big/\spd, \psi\bigr]$ with respect to $\psi(\vx)$.

The inhomogeneous version of Percus's relation \eqref{eq:PercusIdea} can be derived along the same line as the derivation of the homogeneous version in Ref.~\onlinecite{Hansen2013} by simply removing the assumption of the homogeneity.
In an inhomogeneous and symmetry-broken phase, however, the naive definition of the distribution functions loses its uniqueness.
In some systems, for example, a ferromagnet, the uniqueness of the order parameter is recovered by applying a field that breaks the symmetry of the original system, and the order parameter without the field can be obtained by taking zero-limit of the symmetry-breaking field followed by the thermodynamic limit.
I do not know whether the same procedure applies to the distribution functions of inhomogeneous fluids, thus cannot provide a rigorous proof of Percus's relation for inhomogeneous cases; however, I postulate the validity of Eq.~\eqref{eq:PercusIdea} in this paper.

\section{Applications to 2D Crystals}\label{sec:3}
The free-energy functional method introduced in Sec. \ref{sec:2} is applicable to any 2D and 3D system that is characterized by its single- and two-particle densities.
In this paper, as test cases, I use 2D crystals such as square, honeycomb, and kagome lattices as target structures.
While the interaction potentials that stabilize these target crystals have already been found in previous studies~\cite{Rechtsman2005,Rechtsman2006,Marcotte2011,Chen2011,Zhang2013,Jain2014} and are currently of little novelty; nevertheless, these crystals still serve as good test cases.

From the variational principle for $\tA\bigl[\tpd\big/\spd, \psi\bigr]$, it follows that the intrinsic chemical potential $\icp_{\fix}(\vx)$ is the solution of the Euler-Lagrange equation $\delta \tA\bigl[\tpd\big/\spd, \psi\bigr]/\delta \psi(\vx) = 0$, i.e.,
\begin{equation}
\frac{\tpd(\vx, \vx')}{\spd(\vx')}
= -\frac{\delta \Omega\bigl[\psi\bigr]}{\delta \psi(\vx)}\biggr|_{\psi=\icp_{\fix}}. \label{eq:ELeq}
\end{equation}
A direct solution of this equation is computationally demanding.
Therefore, in this paper, I use a trial function $v_{\trial}(\vx, \{x_{i}\})$ with a set of variational parameters $\{x_{i}\}$ as the interaction potential.
The function $\ivf(\vx)$ then becomes $\ivf_{\trial}(\vx, \{x_{i}\}) = \mu - v_{\trial}(\vx, \{x_{i}\})$, and the functional $\tA[\tpd\big/\spd, \ivf\bigr]$ is reduced to the function of $\{x_{i}\}$.
The set of parameters $\{x_{i}\}$ that maximizes $\tA\bigl[\tpd\big/\spd, \ivf_{\trial}\bigr]$ defines the required interaction potential.

In the absence of the external field, the equilibrium configuration is determined solely by $\beta v(\vx)$, because the grand partition function \eqref{eq:grand_partition_function} depends on $v(\vx)$ only through $\beta v(\vx)$.
Thus, the optimal interparticle potential is linearly dependent on the temperature.

The grand potential $\Omega[\ivf_{\trial}]$ can be expanded in a functional Taylor expansion.
Since $\ivf_{\trial}(\vx)$ for the typical interparticle interactions diverges around $\vx = 0$ due to the short range repulsion, the activity $z(\vx) = \exp[\beta \ivf_{\trial}(\vx)]/\lambda^{2}$ is easier to handle than $\ivf_{\trial}(\vx)$ itself.
The functional Taylor expansion of the grand potential in powers of the deviation of activity $\Delta z(\vx) = z(\vx) - z_{0}$ is
\begin{align}
 \beta \Omega[\ivf_{\trial}] &= -\sum_{n=0}^{\infty} \frac{1}{n!} \idotsint \notag \\
 &\quad \frac{\delta^{n} \ln\Xi[\icp]}{\delta z(\vx_{1})\delta z(\vx_{2})\dots\delta z(\vx_{n})}\biggr|_{z=z_{0}}
 \prod_{i=1}^{n} \Delta z(\vx_{i})\rmd \vx_{i}, \label{eq:expansion_Omega}
\end{align}
where $z_{0} = e^{\beta \mu}/\lambda^{2}$ is the activity in the absence of the fixed particle.
The functional derivatives of $\ln \Xi$ with respect to $z(\vx)$ can be expressed in terms of the multiparticle distribution functions.~\cite{Caillol2002, Hansen2013}
Substituting \eqref{eq:PercusIdea} and \eqref{eq:expansion_Omega} into \eqref{eq:tA}, we obtain
\begin{align}
 \beta \tA\bigl[\tpd\big/\spd, \ivf_{\trial}\bigr] &= \beta \Omega[\icp] + \beta \int \frac{\tpd(\vx, 0)}{\spd(0)} \ivf_{\trial}(\vx) \rmd \vx \notag \\
 &\quad - \int \spd(\vx) \zeta(\vx) \rmd \vx - \frac{1}{2}\iint \bigl[\tpd(\vx, \vx') \notag \\
 &\quad - \spd(\vx)\spd(\vx')\bigr] \zeta(\vx)\zeta(\vx')\rmd \vx\rmd \vx', \label{eq:main_eq_for_tA}
\end{align}
up to the second order in $\Delta z(\vx)$, where $\zeta(\vx) = \Delta z(\vx)/z_{0} = \exp[-\beta v_{\trial}(\vx)] - 1$.
The approximation of the truncated Taylor series is justified if $\zeta(\vx)$ is small.
Therefore, when $v_{\trial}(\vx)$ has a negative value, this approximation may break down at very low temperature.
Hereafter, I restrict the consideration to a case in which the minimum of $\beta v_{\trial}(\vx)$ is $-1$.

As the trial function $v_{\trial}(\vx)$, I use the LJG potential, which is an isotropic potential constructed by adding a Gaussian function to the Lennard-Jones potential.~\cite{Engel2007}
Here, the trial function is the LJG potential with positive and negative Gaussians:
\begin{align}
 v_{\trial}(\vx) &= \epsilon\Biggl\{\biggl(\frac{1}{\alpha x}\biggr)^{12} - 2\biggl(\frac{1}{\alpha x}\biggr)^{6}
 - c_{1} \exp\biggl[-\frac{(\alpha x - x_{1})^{2}}{2\sigma_{v}^{2}}\biggr]\notag \\
 &\quad + c_{2} \exp\biggl[-\frac{(\alpha x - x_{2})^{2}}{2\sigma_{v}^{2}}\biggr]\biggr\}, \label{eq:LJG}
\end{align}
where $\sigma_{v}$, $c_{1}$ and $c_{2}$ denote the Gaussian functions' width, depth, and height, respectively.
In this paper, I use the parameters $\sigma_{v}=\sqrt{0.02}$, $c_{1}=1$, and $c_{2}=4$.
The variational parameters $x_{i}$ ($i=1, 2$) are used to adjust the positions of the Gaussian functions.
The values of the remaining parameters $\alpha$ and $\epsilon$ are determined numerically so that the first minimum of $v_{\trial}(\vx)$ will be at the nearest neighbor atomic distance and the value of the global minimum will be $-1$.
Considering these criteria for the trial function, the functional $\tA[\tpd\big/\spd, \ivf_{\trial}\bigr]$ is reduced to a function $\tA(x_{1}, x_{2})$.

The atomic positions in the target perfect crystals, $\{\va{i}\}$, are the input for the free-energy functional method.
I use units for which the nearest neighbor atomic distance is unity.
In the present paper, I assume that the $i$th particle fluctuates around $\va{i}$ and is independent of the other particles, and that its fluctuation is given by a Gaussian distribution.
The single- and two-particle distribution functions are then
\begin{align}
 \spd(\vx) &= \sum_{i} f_{1}(\vx - \va{i}), \label{eq:spd}\\
 \tpd(\vx, \vx') &= \sum_{i} f_{1}(\vx - \va{i})\sum_{j\ne i} f_{1}(\vx' - \va{j}),\label{eq:tpd}
\end{align}
where $f_{1}(\vx)$ denotes the Gaussian distribution
\begin{equation}
 f_{1}(\vx) = \frac{1}{\pi \sigma^{2}} \exp\biggl[-\Bigl(\frac{\vx}{\sigma}\Bigr)^{2}\biggr]. \label{eq:f1}
\end{equation}
The resulting interaction potential depends on the standard deviation $\sigma$.
In principle, any value of $\sigma$ is permissible in this method.
However, because $\spd(\vx)$ and $\tpd(\vx, \vx')$ with sharp peaks are not preferable for the accuracy of the numerical integration in \eqref{eq:main_eq_for_tA}, the broadest but physically acceptable distribution $f_{1}(\vx)$ is better.
Therefore, in this paper, $\sigma$ is set to $0.15$, which is the typical value of the Lindemann ratio at crystallization.
Since $\tpd(\vx, 0)/\spd(0)$ remains finite near $\vx = 0$, where $\icp_{\fix}(\vx)$ diverges rapidly, the approximations \eqref{eq:spd} and \eqref{eq:tpd} cause the second term at the right-hand side of \eqref{eq:main_eq_for_tA} to diverge.
This is caused by neglecting the correlation between the particles in the derivation of \eqref{eq:tpd}.
To include the non-negligible particle correlation due to the strong repulsion between the particles separated by short distances, I modified \eqref{eq:tpd} by multiplying it by $\exp\bigl[-\beta\bigl(1/|\vx|^{12} - 1\bigr)\bigr]$ for $|\vx| < 1$.
The modification factor is proportional to the ideal gas density in a repulsive external field $1/|\vx|^{12}$.

One of the particles in the perfect crystal, for example $\va{1}$, is set at the origin of the coordinate system.
In general, $\tpd(\vx, 0)$ depends on the choice of $\va{1}$ since each particle is not necessarily equivalent in the configuration of the target structure.
In such a case, the average values of $\tA$ over all possible choices of $\va{1}$ should be used.
However, this is not necessary for the target crystal structures used in this study.

I have numerically determined $\tA(x_{1}, x_{2})$ according to \eqref{eq:main_eq_for_tA} using MC integration as implemented in the VEGAS algorithm in GNU Scientific Library.~\cite{Galassi2007}

\begin{figure}[tbp]
 \centering
 \includegraphics[width=8.5cm,clip]{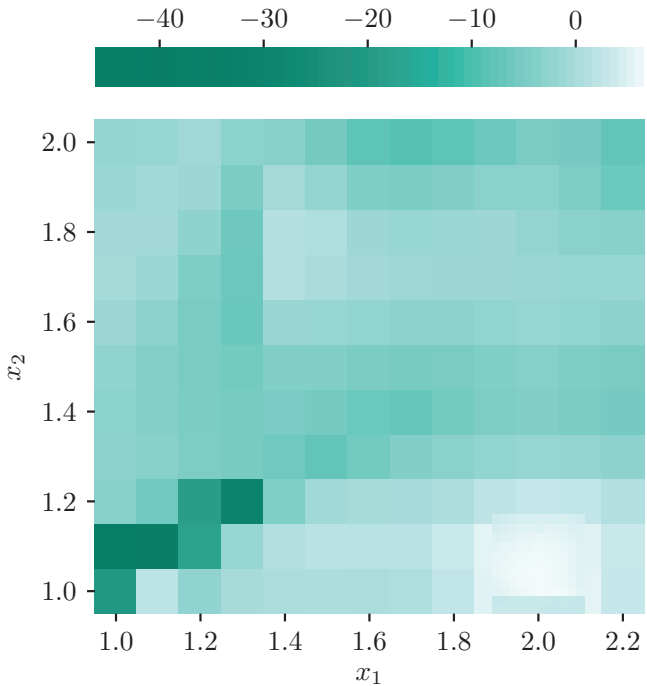}
 \caption{\label{fig:sqa-tA} The density plot of $\tA(x_{1}, x_{2})$ for the square lattice.
 The location of the peak in $\tA(x_{1}, x_{2})$ is first estimated on a coarse grid ($\Delta x_{i} = 0.1$) and is then estimated on a finer grid ($\Delta x_{i} = 0.02$) around the first estimated peak.}
\end{figure}

\begin{figure}[tbp]
 \centering
 \includegraphics[width=8.5cm,clip]{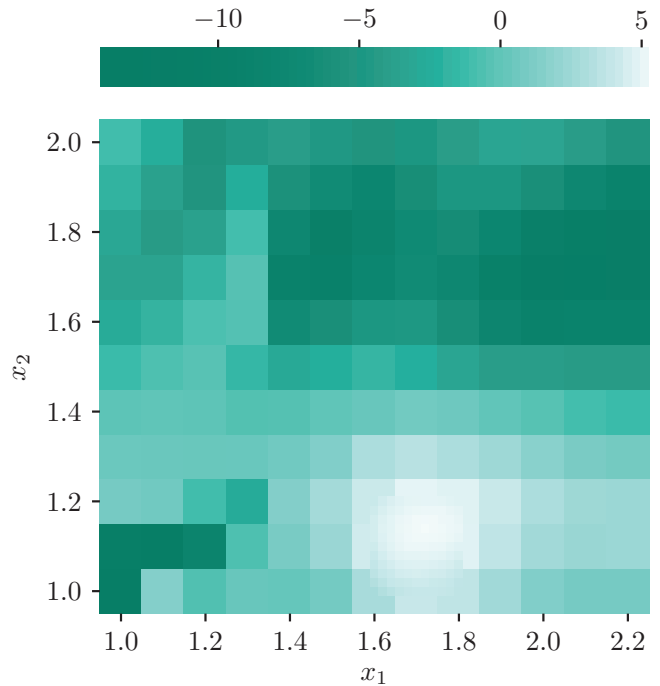}
 \caption{\label{fig:hnc-tA} The density plot of $\tA(x_{1}, x_{2})$ for the honeycomb lattice.
 The estimation grid is the same as in Fig.~\ref{fig:sqa-tA}.}
\end{figure}

\begin{figure}[tbp]
 \centering
 \includegraphics[width=8.5cm,clip]{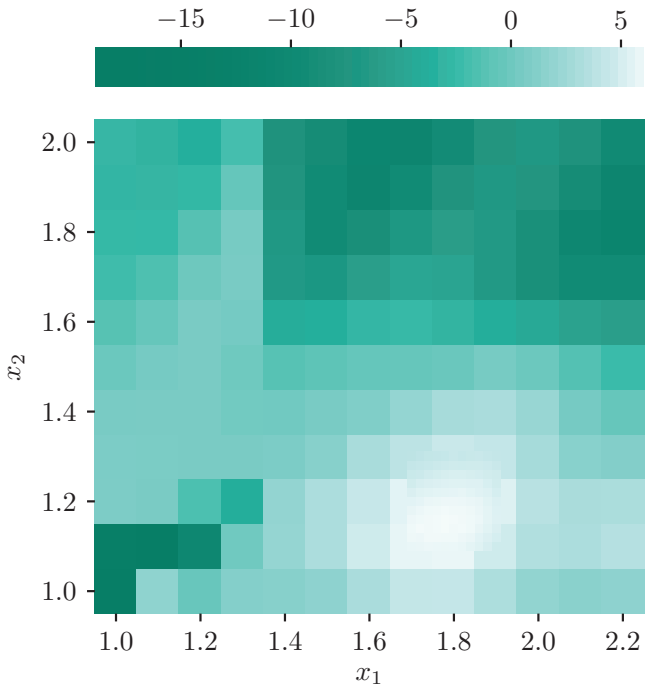}
 \caption{\label{fig:kgm-tA} The density plot of $\tA(x_{1}, x_{2})$ for the kagome lattice.
 The estimation grid is the same as in Fig.~\ref{fig:sqa-tA}.}
\end{figure}

\begin{figure}[tbp]
 \centering
 \includegraphics[width=8.5cm,clip]{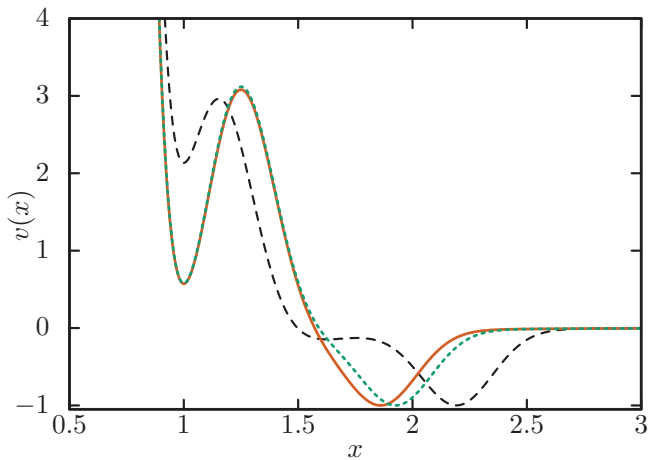}
 \caption{\label{fig:potential} The predicted LJG potentials for square (dashed black curve), honeycomb (solid red curve), and kagome (dotted green curve) lattices.
 These are given by Eq.~\eqref{eq:LJG} with $(x_{1}, x_{2}) = (2.00, 1.04)$, $(1.72, 1.14)$, and $(1.78, 1.14)$, respectively.}
\end{figure}

The density plots of $\tA(x_{1}, x_{2})$ for target lattices are shown in Figs.~\ref{fig:sqa-tA}, \ref{fig:hnc-tA}, and \ref{fig:kgm-tA}.
The peaks in $\tA(x_{1}, x_{2})$ for square, honeycomb, and kagome lattices are located at $(x_{1}, x_{2}) = (2.00, 1.04)$, $(1.72, 1.14)$, and $(1.78, 1.14)$, respectively.
The associated LJG potentials are plotted in Fig.~\ref{fig:potential}.
All predicted LJG potentials have positive values at their first minimum; this reduces the number of the nearest neighbors and prevents the formation of an equilateral triangular lattice.
\begin{table}[tbp]
 \centering
 \caption{\label{table:coordination_number} The radial distances and coordination numbers for the first four nearest neighbors for perfect lattices.}
 \begin{tabular}[t]{l@{\hspace{4ex}}rl@{\hspace{4ex}}rl@{\hspace{4ex}}rl@{\hspace{4ex}}rl}
  \hline
  \hline
         & \multicolumn{2}{c}{Square}& \multicolumn{2}{c}{Honeycomb}& \multicolumn{2}{c}{Kagome}& \multicolumn{2}{c}{Triangular}\\
  \hline
  First  &        $1$,& $4$&        $1$,& $3$&        $1$,& $4$&        $1$,&  $6$\\
  Second & $\sqrt{2}$,& $4$& $\sqrt{3}$,& $6$& $\sqrt{3}$,& $4$& $\sqrt{3}$,&  $6$\\
  Third  &        $2$,& $4$&        $2$,& $3$&        $2$,& $6$&        $2$,&  $6$\\
  Fourth & $\sqrt{5}$,& $8$& $\sqrt{7}$,& $6$& $\sqrt{7}$,& $8$& $\sqrt{7}$,& $12$\\
  \hline
 \end{tabular}
\end{table}
The second minimum of the potential for the square lattice at $x \approx 2.19$ covers both the third and fourth nearest neighbors (see Table~\ref{table:coordination_number} for the neighbor distances and coordination numbers).
Except around the second minimum, the potentials for the honeycomb and kagome lattices show similar behavior.
The fact that these two lattices share the same second and third nearest neighbor distances likely contributes to the similarity in their potentials.
It is reasonable that the second minimum of the honeycomb (kagome) lattice potential is located closer to the second (third) nearest neighbor distance, because the second minimum of the honeycomb (kagome) potential has a larger coordination number at the second (third) nearest neighbor distance than the third (second) one.

The square lattice potential found in Ref.~\onlinecite{Rechtsman2006} has two minima at the first and fourth nearest neighbor distance, like the square lattice potential found here.
But these potentials are not similar, because the former has a deep minimum at the first nearest neighbor distance and a very shallow minimum at the fourth nearest neighbor distance.
The potential for honeycomb lattice here and that in Ref.\onlinecite{Rechtsman2005} have some common features:
the first minimum has positive value and the second minimum is at the second nearest neighbor distance.
The potential for the kagome lattice here is completely different from the known potentials for the kagome lattice, which are purely repulsive.~\cite{Zhang2013,Jain2014}

\begin{figure}[tbp]
 \centering
 \includegraphics[width=8.5cm,clip]{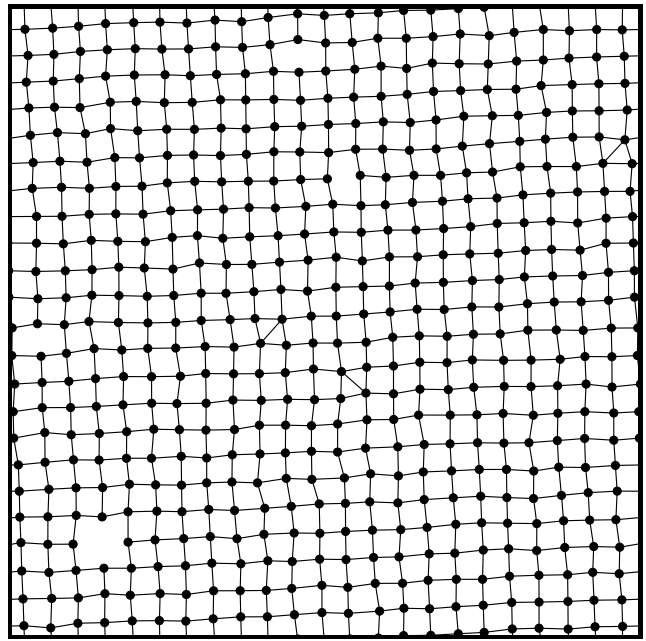}
 \caption{\label{fig:sqa-mc} Results of the 529-particle MC simulations for the LJG potential with $(x_{1}, x_{2}) = (2.00, 1.04)$, annealed from $\kB T = 2.0$ to $\kB T = 1.0$ at $N/V = 1$ in the square simulation box of linear dimension $L=23.0$.
 The particle pairs separated by a distance smaller than $1.2$ are connected by a line segment to guide the eye.}
\end{figure}

\begin{figure}[tbp]
 \centering
 \includegraphics[width=8.5cm,clip]{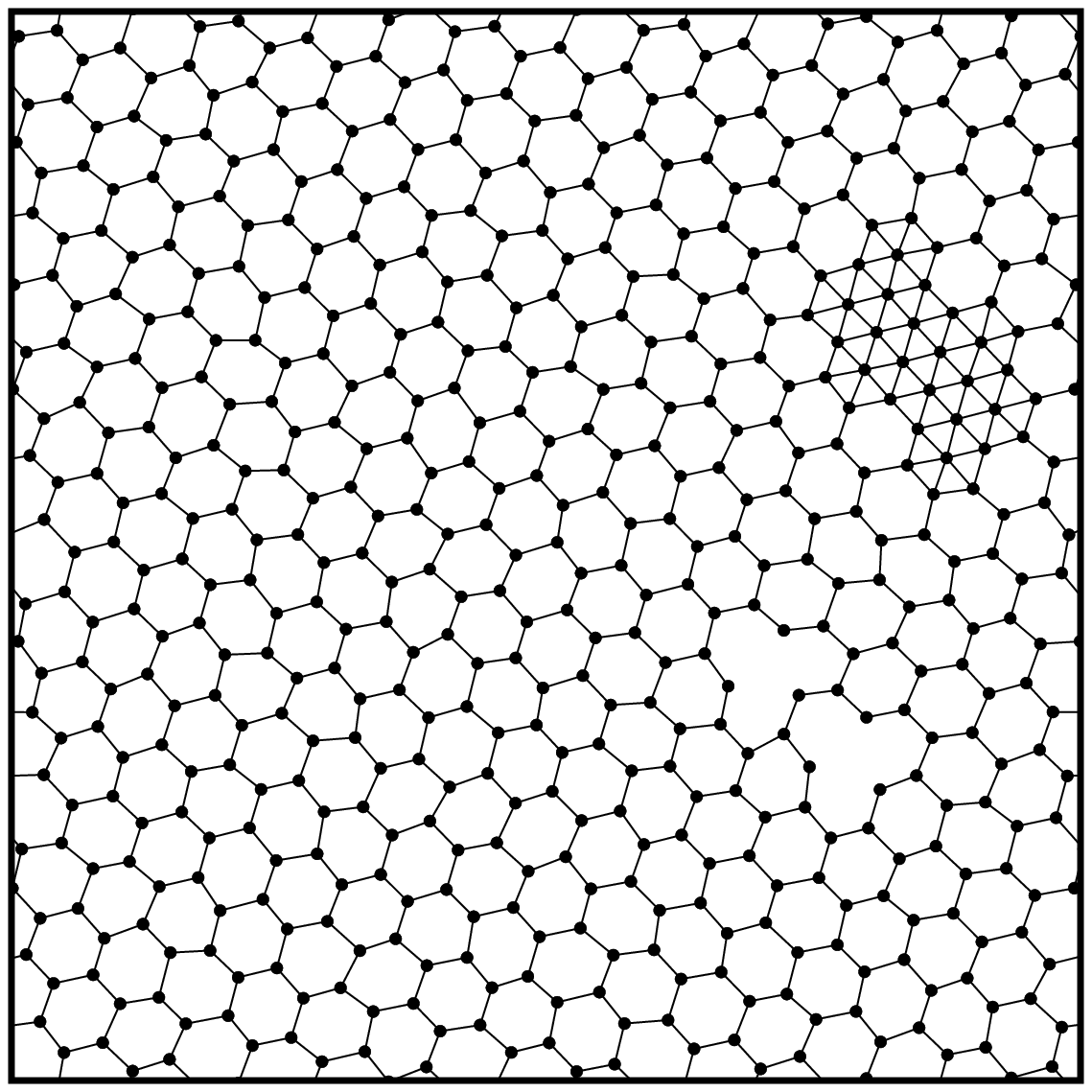}
 \caption{\label{fig:hnc-mc} Results of the 544-particle MC simulations for the LJG potential with $(x_{1}, x_{2}) = (1.72, 1.14)$, annealed from $\kB T = 2.0$ to $\kB T = 0.40$ at $N/V = 4/3\sqrt{3}$ in the square simulation box of linear dimension $L=26.58$.}
\end{figure}

\begin{figure}[tbp]
 \centering
 \includegraphics[width=8.5cm,clip]{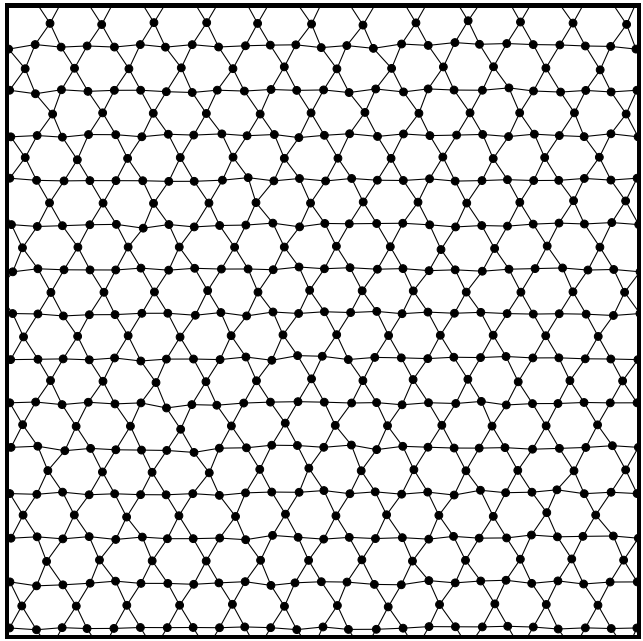}
 \caption{\label{fig:kgm-mc} Results of the 504-particle MC simulations for the LJG potential with $(x_{1}, x_{2}) = (1.78, 1.14)$, annealed from $\kB T = 2.0$ to $\kB T = 0.4$ at $N/V = 3/2\sqrt{3}$ in the square simulation box of linear dimension $L=24.12$.}
\end{figure}

To examine the validity of these predicted potentials, I performed MC simulations for a constant number of particles $N$, volume (area) $V$, and temperature $T$.
For each simulation, $N$ was greater than 500.
The simulation box was a square of side $L$ with periodic boundary conditions;
$L = \sqrt{V}$ was defined so that the density of the system $N/V$ corresponds to that of the target perfect lattice.
If $N$ is not appropriate to fill the simulation box with the unit cells of the target lattice, the success rate of the crystallization into the target lattice will decrease.
For square lattice potential, $N$ was chosen to be a ``magic number,'' with which the unit cells fill the simulation box when one of the sides of the square unit cells and that of the simulation box are parallel.
For the honeycomb and kagome lattice potentials, $N$ was chosen so that the distortion of the lattice remains small when the target lattice fits in to the simulation box under the periodic boundary conditions.
A random configuration was used as an initial configuration for each series of cooling process.
Ten simulation runs were performed for square lattice potential, and all resulting solid phases are square lattice with some defects and no grain boundaries.
For the honeycomb and kagome lattice potentials, most simulation runs resulted in solid phases with grain boundaries and many defects; however, these solid phases exhibited local structures that resembled those of the target lattice.
Several runs (six and four runs, out of thirty runs, for honeycomb and kagome lattice potential, respectively) resulted in crystalline structures without grain boundaries and only a few defects.
The snapshots of these phases are shown in Figs.~\ref{fig:sqa-mc}, \ref{fig:hnc-mc}, and \ref{fig:kgm-mc}, and the figures clearly show that the particles interacting with the LJG potentials predicted by the free-energy functional theory self-assemble into the target crystals.

\section{Discussion and Summary}\label{sec:4}
In this paper, I developed the free-energy functional method for the reconstruction of the interparticle interaction potential from a given structure by combining the variational principle for the functional $\tA[\spd, \psi]$ with Percus's idea.
In this free-energy functional method, the desired interaction potential is given as the function that maximizes $\tA\bigl[\tpd\big/\spd, \psi\bigr]$.
This method was successfully applied to the square, honeycomb, and kagome lattices.

The free-energy functional method introduced here requires single- and two-particle densities, $\spd(\vx)$ and $\tpd(\vx, \vx')$, as input.
To obtain the interaction potential corresponding to an artificially designed structure, e.g., the sets of atomic positions $\{\va{i}\}$, $\spd(\vx)$ and $\tpd(\vx, \vx')$ must be designed or approximated as was done in this work.
If, instead, the method is used to obtain a model potential that reproduces an experimentally observed structure, experimental data for $\spd(\vx)$ and $\tpd(\vx, \vx')$ are sufficient.

The self-assembly of the target lattices in the MC simulations shows that the free-energy functional method works properly.
However, this does not necessarily mean that the interparticle interactions obtained here are entirely optimized for the assembly of the target lattices.
Indeed, if we perform several MC simulations using $(x_{1}, x_{2})=(1.70, 1.06)$, which is slightly different from the predicted parameters for the honeycomb lattice, the honeycomb lattice of comparable quality to the one shown in Fig.~\ref{fig:hnc-mc} is obtained more frequently than when the predicted parameter is used.
This is likely due to the simplistic approximations employed here, such as the second order functional Taylor expansion in the activity for $\tA$ in Eq.~\eqref{eq:main_eq_for_tA}, the use of the Gaussian approximation for $\spd(\vx)$ in Eq.~\eqref{eq:spd}, and the independent-fluctuation approximation for $\tpd(\vx, \vx')$ in Eq.~\eqref{eq:tpd}.
More sophisticated approximations will improve the resulting interparticle interaction potentials.
The maximization of $\tA$ in $(x_{1}, x_{2})$ space may also give rise to the unsatisfactory quality of the resulting structures.
The maximization on the full functional space of interparticle interactions is necessary for the best prediction within the framework of the present theory.

The Taylor expansion of $\tA$ not only affects the accuracy of the prediction but also the range of applicability of this method.
As discussed in Sec. \ref{sec:3}, the choice of the temperature and the trial potential is restricted to justify the truncation of the Taylor series.
Therefore, the present form of this method does not provide a unified description of the optimal interaction potential for a given structure over a wide range (including $T=0$) of temperature.
Unfortunately, better approximations for $\tA[\rho, \psi]$ other than the Taylor series are not available yet.

Even within the truncated Taylor series approximation, the restriction on the temperature can be relaxed using non-negative trial potentials.
Although this is an additional strong restriction on the interaction potential, we can expect that non-negative interaction potentials suffice for a wide variety of target structures at finite temperature, considering that such potentials produce various ground state structures.~\cite{Cohn2009,Jain2013a,Jain2013b,Jain2014}

The interaction potential considered here is restricted to the isotropic one for simplicity.
It was found to be sufficient for the self-assembly of the target crystals considered here, as expected from the past work for the ground state.~\cite{Rechtsman2005,Rechtsman2006,Marcotte2011,Zhang2013,Jain2014}
The class of target crystals that cannot be assembled by isotropic particles is not yet clear,~\cite{Zhang2013} but the isotropy-assumption of the potential will break down and must be discarded if the target structure is strongly anisotropic.

Improving the numerical efficiency is necessary for the future work, e.g., the use of the trial potential with many variational parameters;
the inverse problem for the 3D structure, which requires time-consuming six-dimensional numerical integration in Eq.~\eqref{eq:main_eq_for_tA}.
The use of a terraced (discretized) interparticle potential is a possible candidate for the way to reduce the computation time.
The terraced potentials are used in molecular simulations~\cite{Chapela1989,Xu2006,Bannerman2011} and analytical calculations~\cite{Hollingshead2013,Hollingshead2014} as an efficient and satisfactorily accurate way to investigate the many body systems.

The theory introduced here is also applicable to the inverse problem for liquids; in this case, the interparticle interaction is reconstructed from a given pair distribution function $g(\vx, \vx')$ or a radial distribution function $g(r)$.
In fact, the method is more suitable for homogeneous simple liquids than for crystals, because for liquids, the approximations \eqref{eq:spd} and \eqref{eq:tpd} for distribution functions are unnecessary. 
This is because $\spd(\vx)=\rho$ is constant and $\tpd(\vx, \vx')$ is equal to $\rho^{2}g(\vx, \vx')$ in homogeneous simple liquids.
While the inverse problem for liquids has been solved by the reverse MC method,~\cite{McGreevy1988,Lyubartsev1995} our free-energy functional method serves as a new theoretical approach to tackle this problem.

\begin{acknowledgments}
 I would like to thank Professor Akira Yoshimori for helpful discussions.
\end{acknowledgments}

%

\end{document}